\documentclass[showkeys,eqsecnum,twocolumn,showpacs,preprintnumbers,amssymb,aps,superscriptaddress]{revtex4}
\usepackage{graphicx}
\usepackage{dcolumn}
\usepackage{bm}
\usepackage{latexsym,epsfig}

\def\ii{\textrm{i}}
\newcommand\ea{\emph{et al. }}

\newcommand\mn{$-$}

\def\pols{polarizabilities}
\begin{document}

\preprint{\today}

\title{Ab initio determination of polarizabilities and van der Waals coefficients of Li atoms using the relativistic CCSD(T) method}
\vspace{0.5cm}

\author{L. W. Wansbeek\footnote{wansbeek@kvi.nl}}
\affiliation{KVI, Theory Group, University of Groningen, NL-9747 AA Groningen, The Netherlands}%
\author{ B. K. Sahoo}
\affiliation{KVI, Theory Group, University of Groningen, NL-9747 AA Groningen, The Netherlands}%
\author{R. G. E. Timmermans}
\affiliation{KVI, Theory Group, University of Groningen, NL-9747 AA Groningen, The Netherlands}%
\author{B. P. Das}
\affiliation{Non-accelerator Particle Physics Group, Indian Institute of Astrophysics, Bangalore-560034, India}%
\author{D. Mukherjee}
\affiliation{Department of Physical Chemistry, Indian Association for Cultivation of Science,
Kolkata-700032, India}
\date{\today}
\vskip1.0cm

\begin{abstract}
\noindent We report a new technique to determine the van der Waals coefficients of lithium (Li)
atoms based on the relativistic coupled-cluster theory. These quantities are determined using the
imaginary parts of the scalar dipole and quadrupole polarizabilities, which are evaluated using the
approach that we have proposed in \cite{sahoocpl07}. Our procedure is fully \emph{ab initio}, and
avoids the sum-over-the-states method. We present the dipole and quadrupole polarizabilities of
many of the low-lying excited states of Li. Also, the off-diagonal dipole and quadrupole
polarizabilites between some of the low-lying states of Li are calculated.
\end{abstract}

    \pacs{31.15.Ar,31.15.Dv,31.25.Jf,32.10.Dk}
\keywords{Ab initio method, polarizability}

\maketitle

\section{Introduction}
\noindent In recent years, ultra-cold atom experiments have been used in the study of a variety of
scattering physics, including the probing of different types of phase transitions
\cite{greinernature02}. From an experimental point of view, lithium (Li) is a very interesting
system since its $^6$Li and $^7$Li isotopes correspond to fermionic and bosonic systems,
respectively. These isotopes are used in the study of boson-boson
\cite{moerdijkprl94,quemenerpra07}, boson-fermion \cite{bourdelprl04} and fermion-fermion mixtures
\cite{taglieberprl08,quemenerpra07}.

For the theoretical description of these kinds of systems, a knowledge of the interatomic potential
is necessary. At a large nuclear separation $R$, the $s$-wave scattering interatomic potential is
accurately represented by the sum of two independent contributions, the exchange and electrostatic
potential \cite{smirnovspjetp65}. The former is related to the ionization energies and scattering
lengths which will not be discussed hereafter. The electrostatic potential is given by \cite{yan}
as
\begin{eqnarray}
V(R) = - \frac{C_6}{R^6} - \frac{C_8}{R^8} + \cdots, \label{eq:vdw}
\end{eqnarray}
where $C_6$ and $C_8$ are known as dispersion or van der Waals coefficients. As $R \rightarrow
\infty$, the long-range potential $V(R)$ is dominated by $- C_6/R^6$ and $-C_8/R^8$, where the
higher-order terms are sufficiently weak to be neglected. Both coefficients can be evaluated from
the knowledge of the imaginary parts of the dynamic dipole and quadrupole polarizabilities
\cite{dalgarnoaamp66,dalgarnoacp67}. Several groups have evaluated these quantities because of
their necessity in the simulation, prediction, and interpretation of experiments on optical
lattices in cold-atom collisions, photo-association, and fluorescence spectroscopy
\cite{boestenpra96,amiotpra02}.

Since the classic work of Dalgarno and Lewis \cite{dalgarno55}, different procedures have been
followed to determine polarizabilities. An often-used method is the sum-over-intermediate-states
approach, which employs dipole/quadrupole matrix elements and excitation energies of important
states \cite{dattajpc95,kunducpl91,spelsbergjcp93,sahoorad07,mitroy}. Sum-over-the-states methods
are, however, limited in their accuracy because of the restrictions in the inclusion of higher
states, which are difficult to generate. Coupled-cluster based linear response theory
\cite{dattajpc95,kunducpl91} seems to be a promising method to study both static and dynamic
polarizabilities, while avoiding this limitation of the sum-over-the-states approach. This method
is well applicable to closed-shell systems. For relativistic open-shell systems and adopting atomic
symmetry properties, however, it is not an easy formalism. Therefore, the sum-over-the-states
approach using dipole/quadrupole matrix elements or oscillator strengths is often used in
open-shell atomic systems \cite{sahoorad07,mitroy}.

In this work, we present a novel approach to determine the van der Waals coefficients for lithium
using a method which employs fully atomic symmetry properties in the framework of the relativistic
coupled-cluster (RCC) approach. The approach is \emph{ab initio} and avoids the limitations of the
sum-over-the-states methods. It has recently been employed to determine static polarizabilities in
closed-shell and one-valence open-shell systems \cite{sahoopra08,sahoocpl07,sahoojcmse07}. We also
present the static dipole and quadrupole polarizabilities of many of the excited states in Li.
These could be useful in the calculation of the dispersion coefficients of the excited states and
in determining Stark shifts. So far, only a few studies have been carried out on the
polarizabilities of the Li excited states
\cite{pipinpra92,themelispra95,merawajcp98,zhangpra07,magnierjqsr02,ashbyjqsrt02}. Most of these
studies use non-relativistic theories, and we will compare those results to our relativistic
calculations to assess the relevance of relativistic effects. We also present the scalar
polarizabilities among two different states, which are of interest for several types of studies
\cite{safronovapra99}.

The outline of the rest of the paper is as follows. We start by presenting the theory for
polarizabilities and van der Waals coefficients in Sec. II. Next, we discuss our method of
calculation in Sec. III, and in Sec. IV we present and discuss our results.

\section{Theory}
In this section we give the definitions of the static and dynamic polarizabilities and the van der
Waals coefficients.

\subsection{Polarizability}
\noindent The static dipole polarizability $\alpha_1(J_v,m_{J_v})$ of a valence ($v$) state $|
\Psi_v \rangle$ of a single valence system is given by \cite{angel68, itano00}
\begin{eqnarray}
\alpha_1(J_v,m_{J_v}) &=& \sum_{I \neq v}  \frac { \langle \Psi_v | D | \Psi_I \rangle \langle \Psi_I | D | \Psi_v \rangle } { E_I - E_v} \nonumber \\
 &=& \alpha_1^0(J_v) +\frac{3m_{J_v}^2-J_v(J_v+1)}{J_v(2J_v-1)}\alpha_1^2(J_v), \nonumber\\ \label{eqn1}
 \end{eqnarray}
where the scalar polarizability $\alpha_1^0(J_v)$ is given by
\begin{eqnarray} \alpha_1^0(J_v) &=& \frac{2}{3( 2J_v+1)} \sum_{I \neq v} \frac {
|\langle J_I || D ||J_v \rangle |^2}{E_I - E_v},
\label{eqn2}
 \end{eqnarray}
and the tensor polarizability $\alpha_1^2(J_v)$ by
\begin{eqnarray}
\alpha_1^2(J_v) &=& 2 \left[\frac{10 J_v (2J_v-1)}{3(J_v+1)(2J_v+1)(2J_v+3)}\right]^{1/2} \nonumber
\\&&\sum_{I \neq v} (-)^{J_v-J_I} \left \{ \matrix {J_v & 1 & J_I \cr 1 & J_v & 2 } \right \} \frac
{|\langle J_v || D || J_I \rangle|^2}{E_v - E_I}.\nonumber\\
\label{eqn3}
\end{eqnarray}
Here $D$ is the dipole operator, $J_v$ and $m_{J_v}$ are the angular momentum quantum numbers of $|
\Psi_v \rangle$. $|\Psi_I\rangle$ represents allowed intermediate states with respect to $| \Psi_v
\rangle$ with $E_I$ and $E_v$ their respective energies. Similarly, the scalar quadrupole
polarizability of the valence state $| \Psi_v \rangle$ is given by
\begin{eqnarray}
\alpha_2^0(J_v) &=& \sum_{I \neq v}  \frac { |\langle \Psi_v | Q | \Psi_I \rangle|^2 } { E_I - E_v} \nonumber \\
 &=& \frac{2}{5(2J_v+1)} \sum_{J_I \ne J_v} \frac {|\langle J_v || Q ||J_I \rangle|^2}{E_I - E_v}
 \label{eq:quad}
\end{eqnarray}
where $Q$ is the quadrupole operator.

Extending these definitions, the scalar polarizability between two (possibly different) states
$|\Psi_f\rangle$ and $|\Psi_i\rangle$ is given by \cite{blundellprd92}
\begin{eqnarray}
\alpha_k^0(J_i,J_f) &=& - \sum_{I \neq i,f}\left [ \frac {1}{E_f - E_I} + \frac {1}{E_i -
E_I}\right] \nonumber\\ &&\qquad\times\langle \Psi_f | O^{(k)} | \Psi_I \rangle \langle \Psi_I |
O^{(k)} | \Psi_i \rangle,\nonumber \label{eqn5}
\end{eqnarray}
where  $O^{(k)}$ represents the dipole operator $D$ for $k=1$ and the quadrupole operator $Q$ for
$k=2$, respectively. As a special case the scalar polarizabilities of a state can be recovered by
setting $i=f$ in the above equation. Apart from the static polarizability, a dynamic polarizability
can also be defined. The imaginary part of the dynamic polarizability between two states is given
by
\begin{eqnarray}
\alpha_k^0(\ii \omega) &=& \nonumber- \sum_{I \neq i,f}\left [ \frac {E_f - E_I}{(E_f - E_I)^2 +
\omega^2} + \frac {E_i - E_I}{(E_i - E_I)^2 + \omega^2}\right] \\
&&\qquad\times\langle \Psi_f | O^{(k)} | \Psi_I \rangle \langle \Psi_I | O^{(k)} | \Psi_i \rangle ,
\label{eqn5}
\end{eqnarray}
where $\omega$ is the frequency of the external electromagnetic field.

From these definitions it follows that the determination of the polarizabilities requires the
evaluation of transition matrix elements and the excitation energies, hence a powerful many-body
approach is necessary to evaluate the above quantities to high accuracy.

\subsection{Van der Waals coefficients}
\noindent The general expression for the van der Waals coefficients between two different atoms $a$
and $b$ in terms of their dynamic polarizabilities is given by \cite{dalgarnoaamp66}
\begin{eqnarray}
C^{ab}_{2n} = \label{eq:ccoef}\frac{(2n-2)!}{2
\pi}\sum_{l=1}^{n-2}\frac{1}{(2l)!(2l')!}\int\limits_0^\infty\alpha^a_l(\textrm{i}\omega)\alpha^b_{l'}(\textrm{i}\omega)\textrm{d}\omega,
\end{eqnarray}
where $l' \equiv n-l-1$ and $\alpha^a_l(\textrm{i}\omega)$ and $\alpha^b_{l'}(\textrm{i}\omega)$
are the $2^l$-pole polarizability of atom $a$ and $2^{l'}$-pole polarizability of atom $b$,
respectively. In this article, we evaluate the $C_6$ and $C_8$ coefficients for the $s$-wave ground
state of the Li atom using the simple formulas
\begin{eqnarray}
C_6 &=& \frac{3}{\pi}\int\limits_0^\infty \textrm{d}\omega [ \alpha_1(\textrm{i}\omega)]^2 \label{eq:c6},\\
C_8 &=& \frac{15}{\pi}\int\limits_0^\infty \textrm{d}\omega [
\alpha_1(\textrm{i}\omega)\alpha_2(\textrm{i}\omega) ], \label{eq:c8}
\end{eqnarray}
obtained from Eq. (\ref{eq:ccoef}). The long-range part of the interaction between \emph{three}
ground-state atoms is not exactly equal to the interaction energies taken in pairs.  There is an
extra term which comes from the third-order perturbation. This correction to the van der Waals
potential can be given as $V(R) \propto -{v}/{R^3}$, where \cite{yan}
\begin{eqnarray}
v=\frac{3}{\pi}\int\limits_0^\infty \textrm{d}\omega [\alpha_1(\textrm{i}\omega)]^3, \label{eq:v}
\end{eqnarray}
is called the triple-dipole constant. We have also determined this quantity $v$ for the Li atom and
present the result here.

\section{Method of Calculation}
\noindent The aim of this work is to evaluate Eq. (\ref{eqn5}) for both static ($\omega=0$) and
dynamic (finite $\omega$) polarizabilities, while avoiding the sum-over-intermediate-states
approach and at the same time treating electron-correlation effects rigourously. Coupled-cluster
(CC) theory is one of the most powerful methods to incorporate the electron-correlation effects to
all orders in the atomic wave functions. We employ here a relativistic CC theory that can determine
the atomic wave functions accurately.

Using Eq. (\ref{eqn5}), we write for the dynamic polarizability between states $|\Psi_f\rangle$ and
$|\Psi_i\rangle$
\begin{eqnarray}
 \alpha_k (\textrm{i} \omega) &=& \langle \Psi_f| O^{(k)}| \Psi_i'\rangle
  + \langle \Psi_f' |O^{(k)}| \Psi_i\rangle.
\label{eqn11}
\end{eqnarray}
Comparing Eq. (\ref{eqn5}) and Eq. (\ref{eqn11}), we can express $|\Psi_v'\rangle$, where $v=i,f$,
as
\begin{eqnarray}
|\Psi_v'\rangle &=& \sum_{I\neq v}
\frac{E_I-E_v}{(E_I-E_v)^2+\omega^2}|\Psi_I\rangle\langle\Psi_I|O^{(k)}|\Psi_v\rangle\nonumber\\
 &=& \frac{H_I-E_v}{(H_I-E_v)^2+\omega^2}\sum_{I\neq v}|\Psi_I\rangle \langle \Psi_I| O^{(k)}|\Psi_v\rangle\nonumber\\
&=&\frac{H-E_v}{H-E_v+\ii \omega}\nonumber\\&&\times\left[|\Psi_v\rangle\langle\Psi_v|+\sum_{I \neq
v}|\Psi_I\rangle\langle\Psi_I|
\right]O^{(k)}|\Psi_v\rangle\nonumber\\
&=&\frac{H-E_v}{H-E_v+\ii \omega}\, O^{(k)}|\Psi_v\rangle\nonumber\\
&=&\left[\frac{1}{H-E_v-\ii \omega}\right]\left[\frac{H-E_v}{H-E_v+\ii
\omega}\,O^{(k)}\right]|\Psi_v\rangle,\nonumber\\ \label{eq:3.2}
\end{eqnarray}
where $H$ is the Dirac-Coulomb Hamiltonian. If we next define an effective Hamiltonian
\[
H_\textrm{\scriptsize eff}= H-\ii\omega
\]
and an effective dipole or quadrupole operator
\[
O^{(k)}_\textrm{\scriptsize eff} =\frac{H-E_v}{H-E_v+\ii \omega}O^{(k)}
\]
we can find $|\Psi_v'\rangle$ as the solution of
\begin{eqnarray}
(H_\textrm{\scriptsize eff}-E_v)  |\Psi_v'\rangle=O^{(k)}_\textrm{\scriptsize eff}|\Psi_v\rangle,
\label{eq:heff}
\end{eqnarray}
where $|\Psi_v'\rangle$ are the first-order perturbed wave functions due to the external field.

\subsection{Determination of the DC wave functions}
\noindent To carry out our calculations, we will use CC cluster theory. As this has been described
in detail in many other papers , we will limit ourselves to a short overview. In CC theory, the
atomic wave function $|\Psi_v \rangle$ due to the real part of the effective Hamiltonian of a
single valence ($v$) open-shell system can be expressed as \cite{mukherjee77,lindgren78,lindgren85}
\begin{eqnarray}
|\Psi_v\rangle = \textrm{e}^T \{1+S_v \} |\Phi_v \rangle , \label{eqn14}
\end{eqnarray}
where we define the reference state $|\Phi_v \rangle= a_v^{\dagger}|\Phi_0\rangle$, with
$|\Phi_0\rangle$ the closed-shell Dirac-Fock (DF) state, which is taken as the Fermi vacuum. $T$
and $S_v$ are the CC excitation operators for core to virtual electrons, and valence-core to
virtual electrons, respectively. The curly bracket in the above expression represents the
normal-ordered form. In our calculation, we consider all possible single (S) and double (D)
excitations, as well as the most important triple (T) excitations, an approximation known as the
CCSD(T) method \cite{kaldorjcp87}. To determine the amplitudes of the CC excitation operators we
use

\begin{eqnarray}
\langle \Phi^L | H_c |\Phi_0 \rangle &=& \Delta E_0 \ \delta_{L,0}, \nonumber \\
\langle \Phi_v^K|H_c S_v|\Phi_v\rangle
&=&- \langle \Phi_v^K|H_c |\Phi_v\rangle \nonumber\\&&+ \langle \Phi^K_v|S_v|\Phi_v\rangle \langle \Phi_v|H_c \{1+S_v\} |\Phi_v\rangle\delta_{K,0} \nonumber \\
 &=& - \langle \Phi_v^K|H_c|\Phi_v\rangle +\langle \Phi^K_v|S_v|\Phi_v\rangle \Delta E_v\delta_{K,0}\nonumber\\\label{eqn:15a}
\end{eqnarray}
where we have defined $H_c \equiv \left \{ H_N\textrm{e}^T \right \}_c$. The superscript $L\, (=
1,2)$ represents the singly or doubly excited states from the closed-shell reference (DF) wave
function and $\Delta E_0$ is the correlation energy for the closed-shell system. Further, $\Delta
E_v$ is the electron affinity energy of the valence electron $v$, $K\, (= 1,2)$ denotes the singly
or doubly excited states from the single valence reference state, and the subscripts $N$ and $c$
represent the normal-ordered form and connected terms, respectively. Eqs. (\ref{eqn:15a}) are
non-linear, and they are solved self-consistently by using a Jacobi iterative procedure. With the
amplitudes of the CC excitation operators known, the zeroth-order wave functions can be calculated
by using Eq. (\ref{eqn14}).

\begin{table*}[t]
\caption{The static dipole polarizability $\alpha_1$ of many low-lying levels in Li [au].}
\label{tab:dippol}
\begin{ruledtabular}
\begin{center}
\begin{tabular}{lcccccc}
Level&\multicolumn{2}{c}{\emph{Experiments}}&\multicolumn{2}{c}{\emph{Other theoretical works}} & \multicolumn{2}{c}{\emph{This work}} \\
 &  Scalar&Tensor & Scalar & Tensor & Scalar & Tensor \\
\hline
& 164(3.4)$^a$ &&162.3$^e$,  &   &  & \\
\raisebox{1ex}[0pt]{2s $^2S_{1/2}$}& 164.2(1.1)$^b$ &&164$^f$&&\raisebox{1ex}[0pt]{162.87}\\
&  & &4133$^e$, 4098$^f$ & &\\
\raisebox{1ex}[0pt]{3s $^2S_{1/2}$ } &\raisebox{1.5ex}[0pt]{-}&4136$^c$ & 3832$^d$ & &\raisebox{1.5ex}[0pt]{4107}\\
&&&3.526$\times 10^4$$^e$,\\
\raisebox{1ex}[0pt]{4s $^2S_{1/2}$} &  - && 35040$^f$ & & \raisebox{1ex}[0pt]{3.449 $\times 10^4$ }& \\
&127(3.4)$^i$\\
\raisebox{1ex}[0pt]{2p $^2P_{1/2}$} & 126.9(6)$^g$& & \raisebox{1ex}[0pt]{117.8$^e$} & &\raisebox{1ex}[0pt]{129.41} & \\
3p $^2P_{1/2}$ & - && 2.835$\times 10^4$$^e$  & & 2.938 $\times 10^4$  & \\
4p $^2P_{1/2}$ & - && 2.734$\times 10^5$$^e$  & & 2.635 $\times 10^5$  & \\
\raisebox{0ex}[0pt]{2p $^2P_{3/2}$} & 127.2(7)$^g$ &\raisebox{0ex}[0pt]{1.64(4)$^g$} &\raisebox{0ex}[0pt]{ 117.8$^e$} & \raisebox{0ex}[0pt]{3.874$^e$} & \raisebox{0ex}[0pt]{123.09} &\raisebox{0ex}[0pt]{ 5.95 }\\
3p $^2P_{3/2}$ & -  &-& 2.835$\times 10^4$$^e$ & $-$2173$^e$ & 2.929 $\times 10^4$  & $-$2078   \\
4p $^2P_{3/2}$ & -  &-& 2.735$\times 10^5$$^e$ & $-$2.074$\times 10^4$$^e$ & 2.634 $\times 10^5$  & $-$1.473 $\times 10^4$  \\
3d $^2D_{3/2}$ & $-$15130(40)$^h$ &1.643(6)$\times 10^4$$^h$& $-$1.504$\times 10^4$$^e$ & 1.147$\times 10^4$$^e$ & $-$1.953 $\times 10^4$  & 1.412 $\times 10^4$  \\
4d $^2D_{3/2}$ & -  &-& 3.093$\times 10^6$$^e$ &  $-$5.355$\times 10^5$$^e$ & 3.834 $\times 10^6$  & $-$6.650 $\times 10^5$  \\
3d $^2D_{5/2}$ & $-$15130(40)$^h$ &-& $-$1.510$\times 10^4$$^e$ & 1.645$\times 10^4$$^e$ & $-$2.008 $\times 10^4$  & 2.139 $\times 10^4$  \\
4d $^2D_{5/2}$ & - & -& 3.103$\times 10^6$$^e$ & $-$7.678$\times 10^5$$^e$& 3.843 $\times 10^6$  & $-$9.496 $\times 10^5$ \\
\end{tabular}
$^a$ Molof \emph{et al.} (1974) \cite{molof}, $^b$ Miffre \emph{et al.} (2006) \cite{miffre}, $^c$
Themelis \emph{et al.} (1995) \cite{themelispra95}, $^d$ M\'erawa \emph{et al.} (1998)
\cite{merawajcp98}\\$^e$ Ashby \ea (2003) \cite{ashbyjqsr03}, $^f$ Magnier \ea (2002)
\cite{magnierjqsr02}, $^g$ Windholz \ea (1992) \cite{windholzpra02} ($^6$Li values), $^h$ Ashby \ea
(2003) \cite{ashbyepjd03},\\ $^i$ Hunter \ea (1991) \cite{hunterpra91}.
\end{center}
\end{ruledtabular}
\label{tab1}
\end{table*}
\subsection{Determination of the first-order wave functions}\noindent
\noindent The next step is to determine the first-order wave functions. We write the wave function
of a state with valence electron $v$ in the presence of an external field as
\begin{eqnarray}
|\tilde{\Psi}_v \rangle &=& |\Psi_v\rangle + |\Psi_v'\rangle, \label{eqn16}
\end{eqnarray}
where $|\Psi_v\rangle$ is the wave function of the system in the absence of the external field and
$|\Psi'_v\rangle$ is the first-order correction to $|\Psi_v\rangle$ due to the external field. In
the spirit of the CC approach, we take the ansatz
\begin{eqnarray}
|\tilde{\Psi}_v \rangle = \textrm{e}^{\tilde{T}} \{1+\tilde{S}_v \} |\Phi_v \rangle , \label{eqn17}
\end{eqnarray}
where $\tilde{T}$ and $\tilde{S}_v$ are defined as
\begin{eqnarray}
\tilde{T} &=& T + T', \label{eq:T}\\
\tilde{S}_v &=& S_v + S_v'. \label{eq:S}
\end{eqnarray}
Here $T'$ and $S_v'$ are the corrections to the $T$ and $S_v$ operators in the
presence of the operator $O^{(k)}_\textrm{\scriptsize eff}$, respectively.

Substituting Eqs. (\ref{eq:S}) and (\ref{eq:T}) in Eq. (\ref{eqn17}), we find
\begin{eqnarray}
|\tilde{\Psi}_v \rangle  &=& \textrm{e}^T [1+S_v+ T' \{1+S_v \} + S_v' ] |\Phi_v \rangle,\nonumber\\
\label{eqn18}
\end{eqnarray}
where only the terms linear in $T'$ and $S_v'$ exist, since Eq. (\ref{eq:heff}) contains just one
$O^{(k)}_\textrm{\scriptsize eff}$ operator. By comparing Eqs. (\ref{eqn14}), (\ref{eqn16}), and
(\ref{eqn18}), we get
\begin{eqnarray}
|\Psi_v' \rangle &=& \textrm{e}^T  [T' \{1+S_v \} + S_v' ] |\Phi_v \rangle . \label{eqn19}
\end{eqnarray}
We evaluate these perturbed CC operator amplitudes using the following equations ({\it cf.} Eqs.
(\ref{eqn:15a})):
\[
\langle \Phi^L | \left[H_c^2 + \omega^2\right]T' |\Phi_0 \rangle =  \langle \Phi^L | \left \{
O^{(k)} \textrm{e}^T \right \}_c |\Phi_0 \rangle,\nonumber
\]
\begin{eqnarray}
\langle \Phi_v^K|\left[(H_c\right.&-&\left.\Delta E_v)^2 +\omega^2\right]S_v' |\Phi_v\rangle\nonumber\\
&=& \langle \Phi_v^K | \left \{ O^{(k)}\textrm{e}^T \right \}_c |\Phi_v \rangle \nonumber
\\&-&\langle \Phi_v^K
|\left(H_c+\frac{\omega^2}{H_c}\right)T^{(1)}\{1+S_v^{(0)}\}|\Phi_v\rangle\nonumber\\\label{eqn20}
\end{eqnarray}
where the meaning of $L$ and $K$ was explained above. The first-order wave functions are determined
using Eq. (\ref{eqn19}) after obtaining the perturbed CC amplitudes.

\subsection{Evaluation of $\alpha$ using the RCC approach}\noindent
The expression for the polarizabilities using our CC approach can now be obtained by substituting
Eqs. (\ref{eqn14}) and (\ref{eqn19}) in Eq. (\ref{eqn11}). In this way we get (we also normalize)
\begin{eqnarray}
\alpha_k (\textrm{i} \omega)&=&\frac{\langle \Psi_f| O^{(k)} | \Psi_i'\rangle
+\langle \Psi_f' |O^{(k)}| \Psi_i\rangle }{\sqrt{\langle \Psi_f| \Psi_f\rangle \langle \Psi_i| \Psi_i\rangle}}\nonumber\\
&=&\frac{1}{\sqrt{N_iN_f}}\times\nonumber\\&&\left(\langle\Phi_f|\{1+S_f^\dagger\}\overline{O^{(k)}}\large[T'\{1+S_i\}+S_i'\large]|
\Phi_i\rangle+\right.\nonumber\\
&+&\left.\langle\Phi_f|[S_f'^\dagger+\{1+S_f^\dagger\}T'^\dagger]\overline{O^{(k)}}
\{1+S_i\}|\Phi_i\rangle\right)\nonumber\\
\end{eqnarray}
where
\[
N_v = \langle\Phi_v|\{1+S_v^\dagger\}\mathcal{N}_0\{1+S_v\}|\Phi_v\rangle,
\]
with $v = i, f$, and we have defined $\overline{O^{(k)}} = \textrm{e}^{T^{\dagger}}
O^{(k)}\textrm{e}^T$ and $\mathcal{N}_0=\textrm{e}^{T^{\dagger}} \textrm{e}^T$.

We first evaluate, by using the generalized Wick's theorem, the intermediate terms
$\overline{O^{(k)}}$ and $\mathcal{N}_0$ in the above expressions as effective one-body, two-body,
and so on, terms. Next we sandwich the open-shell valence-core electron excitation operators to
evaluate the exact expression.

\section{Results and Discussions}\noindent We have used partly numerical and partly analytical orbitals to generate the complete
basis sets. The numerical orbitals were obtained using GRASP \cite{parpia}, and the analytical
orbitals were obtained using Gaussian-type orbitals (GTO's) \cite{chaudhuripra99}. In total, we
have taken up to the 30$s$, 30$p$, 25$d$, 25$f$, and 20$g$ orbitals to calculate the DF wave
function. Out of these, we have generated the first 4, 3, 2, 2, and 2 orbitals from the $s$, $p$,
$d$, $f$, and $g$ symmetries, respectively, using GRASP. The remaining continuum orbitals were
obtained analytically from GTO's, using as parameters $\alpha= 0.00525$ and $\beta = 2.73$. After
this, the final orbitals were orthogonalized using Schmidt's procedure \cite{majumderjpb01}.
\begin{figure}[h]
\includegraphics[width=5.0cm,clip=true]{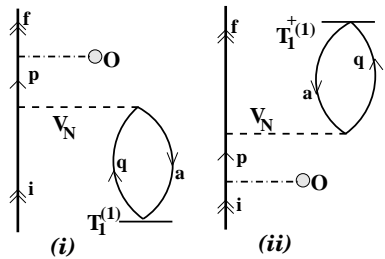}
\caption{Extra correlation diagrams which appear in the calculation of the polarizabilities using
our novel approach. These diagrams do not appear when the CC wave functions are used in the
sum-over-the-states method. } \label{fig1}
\end{figure}

We present the static dipole and quadrupole polarizabilities of several important low-lying states
of Li in Table
 \ref{tab1} and Table \ref{tab2}, respectively. In these Tables, we have
also listed other theoretical results and the most recent experimental results, where available.
For the ground state, a number of theoretical dipole polarizability results are available, for the
excited states, however, few calculations have been carried out. All other theoretical results
except one are based on non-relativistic theory. Some of these calculations are also performed
using molecular codes, at the cost of atomic symmetries \cite{magnierjqsr02}. The one available
relativistic calculation on the excited states is carried out using a rather approximate method to
include the correlation effects due to the Coulomb interaction \cite{ashbyjqsr03}. Our calculation
uses a relativistic approach which considers correlation effects to all orders in the form of CC
amplitudes.
\begin{figure}[h]
\includegraphics[scale=0.4]{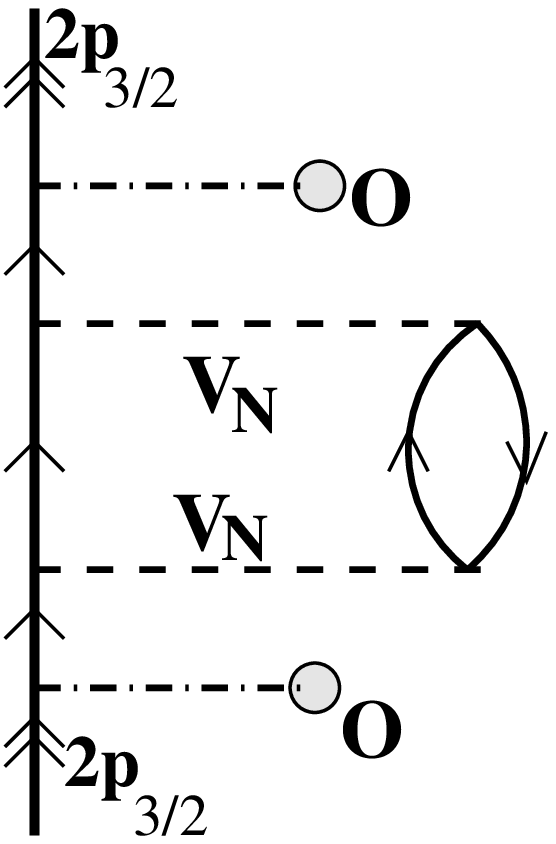}
\caption{The correlation diagram that causes a large discrepancy between the calculated and the
experimental results of the tensor polarizability of the $2p_{3/2}$ state.} \label{fig2}
\end{figure}
Table \ref{tab:quadpol} shows the result for the static quadrupole \pols. No experimental data is
available for comparison, and the available theoretical results for the $2S_{1/2}$ level are not
very consistent.
\begin{table} \caption{The static quadrupole polarizability $\alpha_2^0$ of many important states in
Li [au].}\label{tab:quadpol}
\begin{ruledtabular}
\begin{center}
\begin{tabular}{lcc}
 Level&  \emph{Other theoretical works} & \emph{This work} \\
\hline
 & 1423$^a$, 1424$^b$,1430$^c$, 1423.266(5)$^d$,&   \\
\raisebox{1.5ex}[0pt]{2s $^2S_{1/2}$}&1403$^e$, 1393$^f$, 1424(4)$^g$, 1424.4$^h$  &\raisebox{1.5ex}[0pt]{1420}   \\
3s $^2S_{1/2}$ &3.5642$\times 10^5$$^h$ & 3.475 $\times 10^5$ \\
4s $^2S_{1/2}$ & 1.1587$\times10^7$$^h$ & 1.113$ \times 10^7$ \\
2p $^2P_{1/2}$ &  - & 7.804 $ \times 10^4$  \\
3p $^2P_{1/2}$ &  - & $1.033 \times 10^7$ \\
4p $^2P_{1/2}$ &  - & $3.301 \times 10^9$ \\
\end{tabular}
$^a$ Spelsberg \emph{et al.} (1993) \cite{spelsbergjcp93}, $^b$ Marinescu \emph{et al.} (1994)
\cite{marinescu}, $^c$ M\'erawa \emph{et al.} (1994) \cite{merawa94}, $^d$ Yan \emph{et al.} (1996)
\cite{yan}, $^{e,f}$ Patil and Tang (1997,1999) \cite{patil97,patil99}, $^g$ Snow \emph{et
al.} (2005) \cite{snow05},\\
$^h$ Zhang \ea (2007) \cite{zhangpra07}.
\end{center}
\end{ruledtabular}
\label{tab2}
\end{table}
\begin{table}
\caption{The off-diagonal scalar polarizability in Li [au].}\label{tab:difpol}
\begin{ruledtabular}
\centering
\begin{tabular}{lll}
&DF&CCSD(T)\\
\hline \emph{Dipole}\\
 $2s-3s$&\mn 27.18&\mn 20.41\\
$2s-4s$&\mn 202.9&\mn 164.2\\
$3s-4s$&\mn 105.8&6.292\\
\tabularnewline \emph{Quadrupole}\\
$2s-3s$&2.495 $\times 10^4$&2.219$\times 10^4$\\
$2s-4s$& 1.245 $\times 10^5$& 1.134 $\times 10^5$\\
$3s-4s$&9.281 $\times 10^5$& 6.647 $\times 10^5$\\
\end{tabular}
\end{ruledtabular}
\end{table}

\begin{figure}
\centering
\begin{tabular}{cc}
\includegraphics[scale=0.6]{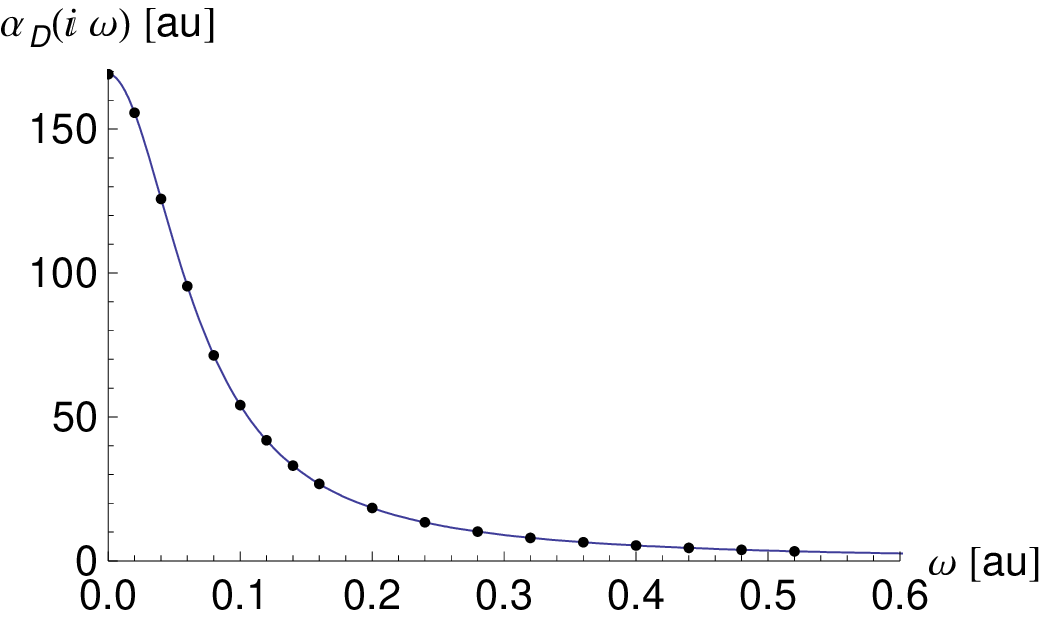}\\
($i$)\\
\includegraphics[scale=0.6]{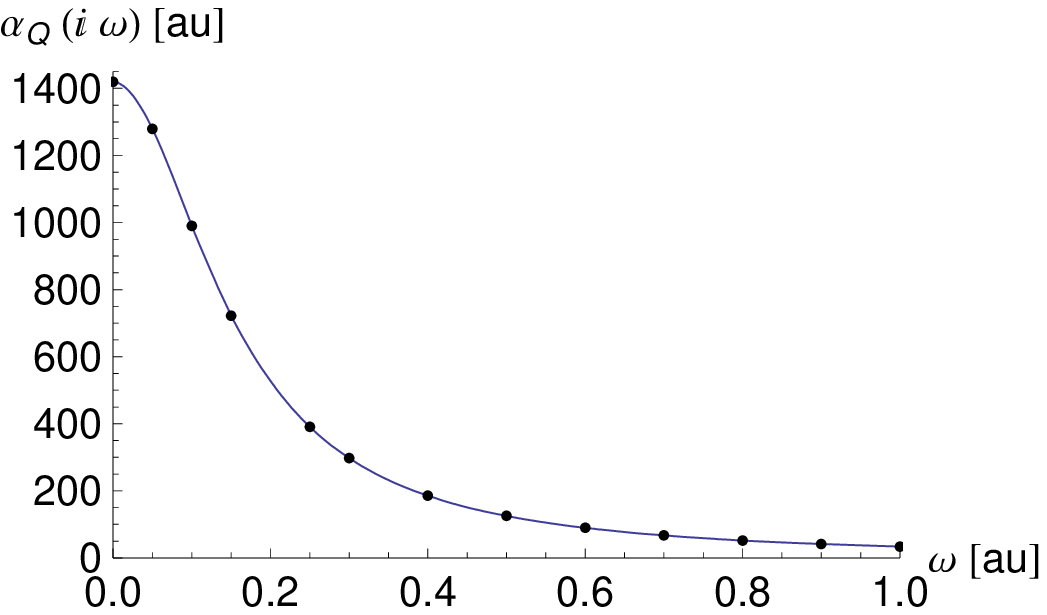}\\
($ii$)
\end{tabular}
\caption{The imaginary parts of the dipole ($i$) and quadrupole ($ii$) polarizabilities  of the
ground state of Li as a function of the angular frequency $\omega$.}\label{fig:dipquad}
\end{figure}
\begin{figure}
\centering
\begin{tabular}{cc}
\includegraphics[scale=0.6]{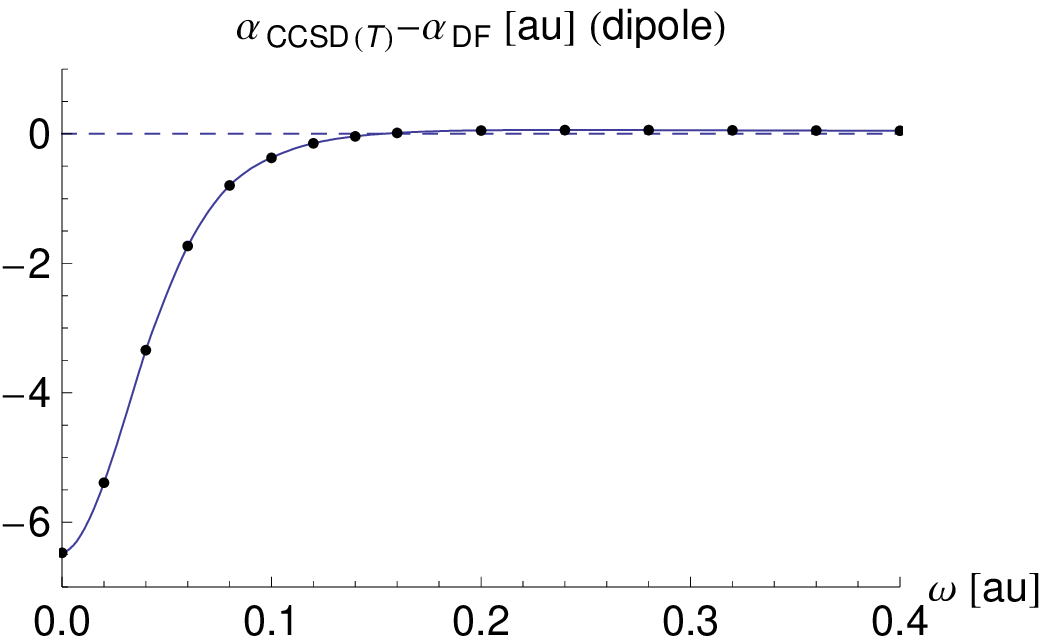}\\
($i$)\\
\includegraphics[scale=0.6]{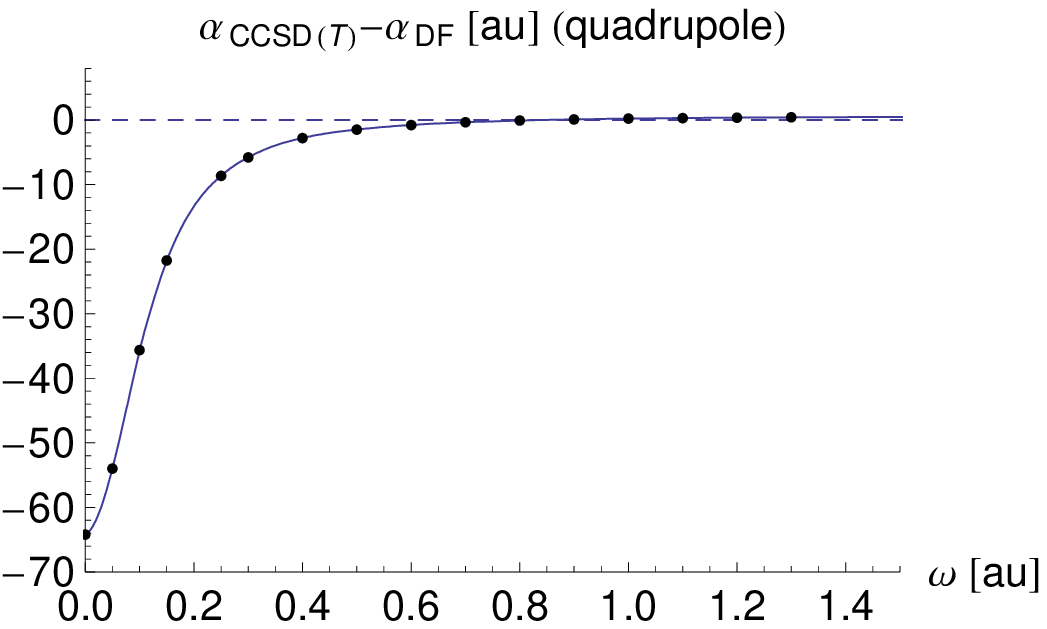}\\
($ii$)\\
\end{tabular}
\caption{The difference between the DF and CCSD(T) results for the imaginary parts of the dipole
($i$) and quadrupole ($ii$) polarizabilities of the ground state of Li as a function of the angular
frequency $\omega$.}\label{fig:dif}
\end{figure}
\begin{table}
  \caption{C$_6$ and C$_8$ values for the ground states of Li-Li [au].}\label{tab:c6c8}
  \begin{ruledtabular}
    \centering
    \begin{tabular}{lll}
    &C$_6$($\times 10^3$)&C$_8$($\times 10^5$)\\
    \hline
    \emph{This work}&&\\
    Dirac-Fock&1.473&0.8891\\
    CCSD(T)&1.396(6)&0.8360\\
    \tabularnewline
    \emph{Other theoretical works}&&\\
    Marinescu \emph{et al.} (1994) \cite{marinescu}&1.388&0.8324\\
    Spelsberg \emph{et al.} (1996)\cite{spelsbergjcp93} &&\\
Yan \emph{et al.} (1996) \cite{yan}&1.39322&0.834258(42)\\
Patil and Tang (1999) \cite{patil99} &1.360&0.8100\\
Porsev and Derevianko (2003) \cite{porsev03} &-&0.834(4)\\
Mitroy and Bromley (2003) \cite{mitroy}&1.3946&0.83515\\
    \end{tabular}
  \end{ruledtabular}
\end{table}

Although our method is theoretically superior to the previously employed methods to determine both
dipole and quadrupole polarizabilities, it seems that some of the earlier results  are in better
agreement with the experimental results than ours. This may be due to the fact that experimental
energies are used in some of these calculations in contrast to our method which is fully {\it ab
initio}. This means that in our calculation there may be strong cancelations with neglected
higher-order excitations in the correlation effects. We note that in our approach we implicitly
take into account certain correlation effects that cannot be accounted for in the usual
sum-over-states approach that is used in so many of the earlier calculations. These diagrams, which
are shown diagrammatically in Fig. \ref{fig1}, are part of the RPA.

As Table \ref{tab:dippol} shows, our value for the tensor polarizability of the $2p^2P_{3/2}$ level
is larger than the experimental result. In our investigation we found that this large value is due
to the unusual behavior of the correlation effects produced by the diagram shown in Fig.
\ref{fig2}. Leaving out this diagrams yields a value for the tensor polarizability of the
$2p^2P_{3/2}$ level of $\sim$ 1.6, which agrees nicely with the experiment. For the completeness of
the theory this effect cannot be left out. We expect that this effect will cancel with the
neglected higher-order excitations.

In Table \ref{tab:difpol}, we present scalar dipole and quadrupole polarizabilities among different
$s$-states of Li which are also important in the determination of the van der Waals coefficients of
the excited states for ultra-cold atom experiments. Our method can also be employed to determine
these quantities in the heavy alkali atoms like Cs and Fr that are important candidates for the
study of atomic parity nonconservation \cite{safronovapra99}. To our knowledge, no other results
are available to compare with these results. As the Table shows, the scalar dipole polarizability
between the $2s$ and $3s$ states in Li is of opposite sign to the other alkali atoms
\cite{safronovapra99}.

\begin{table}
\caption{The triple-dipole constant $v$ for Li-Li-Li ($\times 10^4$) [au].} \label{tab:triple}
  \begin{ruledtabular}
  \centering
    \begin{tabular}{ll}
&$v$(Li-Li-Li)\\
      \hline
      \emph{This work}&\\
      Dirac-Fock&18.576\\
      CCSD(T)&16.934\\
    \tabularnewline
    \emph{Other theoretical works}&\\
    Yan \emph{et al.} (1996) \cite{yan}&17.0595(6)\\
    Mitroy and Bromley (2003) \cite{mitroy}&17.087

    \end{tabular}
  \end{ruledtabular}
\end{table}

The main goal of this work is to illustrate how to evaluate the van der Waals coefficients using
the present method. Figure \ref{fig:dipquad} shows the imaginary parts of the dipole and quadrupole
polarizabilities (in atomic units) of the ground state of Li as functions of angular frequency,
$\omega$. As the Figures show, these quantities fall off exponentially for higher values of
$\omega$. To illustrate the effect of electron correlation as a function of frequency, we have
plotted the difference between the CCSD(T) and the DF results in Fig. \ref{fig:dif}. This Figure
suggests that the correlation effects vanish for higher frequencies. Using the imaginary parts of
the dipole and quadrupole polarizabilities in Eqs. (\ref{eq:c6}), (\ref{eq:c8}) and (\ref{eq:v}),
we evaluated the $C_6$,  $C_8$, and $v$ coefficients, respectively, using a numerical integration
method.

In Table \ref{tab:c6c8} we present our $C_6$ and $C_8$ coefficients and compare them with the other
available results. Although our value for the static polarizability of the ground state of Li is
slightly smaller than the results presented by others, our $C_6$ and $C_8$ values are in good
agreement with the other results. We present the coefficient $v$ of the third-order correction to
the long-range potential in Table \ref{tab:triple}, which matches well with the other available
semi-empirical results.

\section{Conclusion}
\noindent We have employed a novel approach to determine both ground and excited states
polarizabilities by treating the electron-correlation effects and wave functions due to external
operators in the spirit of RCC ansatz. This approach was used to determine the imaginary parts of
the polarizabilities which we used to evaluate the van der Waals coefficients of the Li atom. By
using this novel technique, we were able to consider the electron-correlation effects rigorously
because the technique is fully relativistic and it avoids the sum-over-states method.

\section{Acknowledgment}
\noindent We thank Dr. R. K. Chaudhuri for his contribution in developing some parts of the codes.
We thank the C-DAC TeraFlop Super Computing facility, Bangalore, India for the cooperation to carry
out these calculations on its computers.

\bibliography{cc2}

\begin{thebibliography}{49}
\expandafter\ifx\csname natexlab\endcsname\relax\def\natexlab#1{#1}\fi
\expandafter\ifx\csname bibnamefont\endcsname\relax
  \def\bibnamefont#1{#1}\fi
\expandafter\ifx\csname bibfnamefont\endcsname\relax
  \def\bibfnamefont#1{#1}\fi
\expandafter\ifx\csname citenamefont\endcsname\relax
  \def\citenamefont#1{#1}\fi
\expandafter\ifx\csname url\endcsname\relax
  \def\url#1{\texttt{#1}}\fi
\expandafter\ifx\csname urlprefix\endcsname\relax\def\urlprefix{URL }\fi
\providecommand{\bibinfo}[2]{#2}
\providecommand{\eprint}[2][]{\url{#2}}

\bibitem[{\citenamefont{Sahoo}(2007)}]{sahoocpl07}
\bibinfo{author}{\bibfnamefont{B.~K.} \bibnamefont{Sahoo}},
  \bibinfo{journal}{Chem. Phys. Lett.} \textbf{\bibinfo{volume}{448}},
  \bibinfo{pages}{144} (\bibinfo{year}{2007}).

\bibitem[{\citenamefont{Greiner et~al.}(2002)\citenamefont{Greiner, Mandel,
  Esslinger, H{\"a}nsch, and Bloch}}]{greinernature02}
\bibinfo{author}{\bibfnamefont{M.}~\bibnamefont{Greiner}},
  \bibinfo{author}{\bibfnamefont{O.}~\bibnamefont{Mandel}},
  \bibinfo{author}{\bibfnamefont{T.}~\bibnamefont{Esslinger}},
  \bibinfo{author}{\bibfnamefont{T.~W.} \bibnamefont{H{\"a}nsch}},
  \bibnamefont{and} \bibinfo{author}{\bibfnamefont{I.}~\bibnamefont{Bloch}},
  \bibinfo{journal}{Nature} \textbf{\bibinfo{volume}{415}}, \bibinfo{pages}{39}
  (\bibinfo{year}{2002}).

\bibitem[{\citenamefont{Moerdijk et~al.}(1994)\citenamefont{Moerdijk, Stwalley,
  Hulet, and Verhaar}}]{moerdijkprl94}
\bibinfo{author}{\bibfnamefont{A.~J.} \bibnamefont{Moerdijk}},
  \bibinfo{author}{\bibfnamefont{W.~C.} \bibnamefont{Stwalley}},
  \bibinfo{author}{\bibfnamefont{R.~G.} \bibnamefont{Hulet}}, \bibnamefont{and}
  \bibinfo{author}{\bibfnamefont{B.~J.} \bibnamefont{Verhaar}},
  \bibinfo{journal}{Phys. Rev. Lett.} \textbf{\bibinfo{volume}{72}},
  \bibinfo{pages}{40} (\bibinfo{year}{1994}).

\bibitem[{\citenamefont{Quemener et~al.}(2007)\citenamefont{Quemener, Launay,
  and Honvault}}]{quemenerpra07}
\bibinfo{author}{\bibfnamefont{G.}~\bibnamefont{Quemener}},
  \bibinfo{author}{\bibfnamefont{J.~M.} \bibnamefont{Launay}},
  \bibnamefont{and} \bibinfo{author}{\bibfnamefont{P.}~\bibnamefont{Honvault}},
  \bibinfo{journal}{Phys. Rev. A} \textbf{\bibinfo{volume}{75}},
  \bibinfo{pages}{050701} (\bibinfo{year}{2007}).

\bibitem[{\citenamefont{Bourdel et~al.}(2004)\citenamefont{Bourdel, Khykovich,
  Cubizolles, Zhang, Chevy, Teichmann, Tarruell, Kokkelmans, and
  Salomon}}]{bourdelprl04}
\bibinfo{author}{\bibfnamefont{T.}~\bibnamefont{Bourdel}},
  \bibinfo{author}{\bibfnamefont{L.}~\bibnamefont{Khykovich}},
  \bibinfo{author}{\bibfnamefont{J.}~\bibnamefont{Cubizolles}},
  \bibinfo{author}{\bibfnamefont{J.}~\bibnamefont{Zhang}},
  \bibinfo{author}{\bibfnamefont{F.}~\bibnamefont{Chevy}},
  \bibinfo{author}{\bibfnamefont{M.}~\bibnamefont{Teichmann}},
  \bibinfo{author}{\bibfnamefont{L.}~\bibnamefont{Tarruell}},
  \bibinfo{author}{\bibfnamefont{S.~J.} \bibnamefont{Kokkelmans}},
  \bibnamefont{and} \bibinfo{author}{\bibfnamefont{C.}~\bibnamefont{Salomon}},
  \bibinfo{journal}{Phys. Rev. Lett.} \textbf{\bibinfo{volume}{93}},
  \bibinfo{pages}{050401} (\bibinfo{year}{2004}).

\bibitem[{\citenamefont{Taglieber et~al.}(2008)\citenamefont{Taglieber, Voigt,
  Aoki, H{\"a}nsch, and Dieckmann}}]{taglieberprl08}
\bibinfo{author}{\bibfnamefont{M.}~\bibnamefont{Taglieber}},
  \bibinfo{author}{\bibfnamefont{A.-C.} \bibnamefont{Voigt}},
  \bibinfo{author}{\bibfnamefont{T.}~\bibnamefont{Aoki}},
  \bibinfo{author}{\bibfnamefont{T.~W.} \bibnamefont{H{\"a}nsch}},
  \bibnamefont{and}
  \bibinfo{author}{\bibfnamefont{K.}~\bibnamefont{Dieckmann}},
  \bibinfo{journal}{Phys. Rev. Lett.} \textbf{\bibinfo{volume}{100}},
  \bibinfo{pages}{010401} (\bibinfo{year}{2008}).

\bibitem[{\citenamefont{Smirnov and Chibisov}(1965)}]{smirnovspjetp65}
\bibinfo{author}{\bibfnamefont{B.~M.} \bibnamefont{Smirnov}} \bibnamefont{and}
  \bibinfo{author}{\bibfnamefont{M.~I.} \bibnamefont{Chibisov}},
  \bibinfo{journal}{Sov. Phys. JETP} \textbf{\bibinfo{volume}{21}},
  \bibinfo{pages}{624} (\bibinfo{year}{1965}).

\bibitem[{\citenamefont{Yan et~al.}(1996)\citenamefont{Yan, Babb, Dalgarno, and
  Drake}}]{yan}
\bibinfo{author}{\bibfnamefont{Z.~C.} \bibnamefont{Yan}},
  \bibinfo{author}{\bibfnamefont{J.~F.} \bibnamefont{Babb}},
  \bibinfo{author}{\bibfnamefont{A.}~\bibnamefont{Dalgarno}}, \bibnamefont{and}
  \bibinfo{author}{\bibfnamefont{G.~F.~W.} \bibnamefont{Drake}},
  \bibinfo{journal}{Phys. Rev. A} \textbf{\bibinfo{volume}{54}},
  \bibinfo{pages}{2824} (\bibinfo{year}{1996}).

\bibitem[{\citenamefont{Dalgarno and Davison}(1966)}]{dalgarnoaamp66}
\bibinfo{author}{\bibfnamefont{A.}~\bibnamefont{Dalgarno}} \bibnamefont{and}
  \bibinfo{author}{\bibfnamefont{W.~D.} \bibnamefont{Davison}},
  \bibinfo{journal}{Adv. At. Mol. Phys.} \textbf{\bibinfo{volume}{2}},
  \bibinfo{pages}{1} (\bibinfo{year}{1966}).

\bibitem[{\citenamefont{Dalgarno}(1967)}]{dalgarnoacp67}
\bibinfo{author}{\bibfnamefont{A.}~\bibnamefont{Dalgarno}},
  \bibinfo{journal}{Adv. Chem. Phys.} \textbf{\bibinfo{volume}{12}},
  \bibinfo{pages}{143} (\bibinfo{year}{1967}).

\bibitem[{\citenamefont{Boesten et~al.}(1996)\citenamefont{Boesten, Vogels,
  Tempelaars, and Verhaar}}]{boestenpra96}
\bibinfo{author}{\bibfnamefont{H.~M.} \bibnamefont{Boesten}},
  \bibinfo{author}{\bibfnamefont{J.~M.} \bibnamefont{Vogels}},
  \bibinfo{author}{\bibfnamefont{J.~G.} \bibnamefont{Tempelaars}},
  \bibnamefont{and} \bibinfo{author}{\bibfnamefont{B.~J.}
  \bibnamefont{Verhaar}}, \bibinfo{journal}{Phys. Rev. A}
  \textbf{\bibinfo{volume}{54}} (\bibinfo{year}{1996}).

\bibitem[{\citenamefont{Amiot et~al.}(2002)\citenamefont{Amiot, Dulieu,
  Gutterres, and Masnou-Seeuws}}]{amiotpra02}
\bibinfo{author}{\bibfnamefont{C.}~\bibnamefont{Amiot}},
  \bibinfo{author}{\bibfnamefont{O.}~\bibnamefont{Dulieu}},
  \bibinfo{author}{\bibfnamefont{R.~F.} \bibnamefont{Gutterres}},
  \bibnamefont{and}
  \bibinfo{author}{\bibfnamefont{F.}~\bibnamefont{Masnou-Seeuws}},
  \bibinfo{journal}{Phys. Rev. A} \textbf{\bibinfo{volume}{66}},
  \bibinfo{pages}{052506} (\bibinfo{year}{2002}).

\bibitem[{\citenamefont{Dalgarno and Lewis}(1955)}]{dalgarno55}
\bibinfo{author}{\bibfnamefont{A.}~\bibnamefont{Dalgarno}} \bibnamefont{and}
  \bibinfo{author}{\bibfnamefont{J.~T.} \bibnamefont{Lewis}},
  \bibinfo{journal}{Proc. R. Soc. London} \textbf{\bibinfo{volume}{233}},
  \bibinfo{pages}{70} (\bibinfo{year}{1955}).

\bibitem[{\citenamefont{Datta et~al.}(1995)\citenamefont{Datta, Sen, and
  Mukherjee}}]{dattajpc95}
\bibinfo{author}{\bibfnamefont{B.}~\bibnamefont{Datta}},
  \bibinfo{author}{\bibfnamefont{P.}~\bibnamefont{Sen}}, \bibnamefont{and}
  \bibinfo{author}{\bibfnamefont{D.}~\bibnamefont{Mukherjee}},
  \bibinfo{journal}{J. Phys. Chem.} \textbf{\bibinfo{volume}{99}},
  \bibinfo{pages}{6441} (\bibinfo{year}{1995}).

\bibitem[{\citenamefont{Kundu and Mukherjee}(1991)}]{kunducpl91}
\bibinfo{author}{\bibfnamefont{B.}~\bibnamefont{Kundu}} \bibnamefont{and}
  \bibinfo{author}{\bibfnamefont{D.}~\bibnamefont{Mukherjee}},
  \bibinfo{journal}{Chem. Phys. Lett.} \textbf{\bibinfo{volume}{179}},
  \bibinfo{pages}{468} (\bibinfo{year}{1991}).

\bibitem[{\citenamefont{Spelsberg et~al.}(1993)\citenamefont{Spelsberg, Lorenz,
  and Meyer}}]{spelsbergjcp93}
\bibinfo{author}{\bibfnamefont{D.}~\bibnamefont{Spelsberg}},
  \bibinfo{author}{\bibfnamefont{T.}~\bibnamefont{Lorenz}}, \bibnamefont{and}
  \bibinfo{author}{\bibfnamefont{W.}~\bibnamefont{Meyer}}, \bibinfo{journal}{J.
  Chem. Phys.} \textbf{\bibinfo{volume}{99}}, \bibinfo{pages}{7845}
  (\bibinfo{year}{1993}).

\bibitem[{\citenamefont{Sahoo et~al.}(2007{\natexlab{a}})\citenamefont{Sahoo,
  Das, Chaudhuri, Mukherjee, Timmermans, and Jungmann}}]{sahoorad07}
\bibinfo{author}{\bibfnamefont{B.~K.} \bibnamefont{Sahoo}},
  \bibinfo{author}{\bibfnamefont{B.~P.} \bibnamefont{Das}},
  \bibinfo{author}{\bibfnamefont{R.~K.} \bibnamefont{Chaudhuri}},
  \bibinfo{author}{\bibfnamefont{D.}~\bibnamefont{Mukherjee}},
  \bibinfo{author}{\bibfnamefont{R.~G.~E.} \bibnamefont{Timmermans}},
  \bibnamefont{and} \bibinfo{author}{\bibfnamefont{K.}~\bibnamefont{Jungmann}},
  \bibinfo{journal}{Phys. Rev. A} \textbf{\bibinfo{volume}{76}},
  \bibinfo{pages}{040504(R)} (\bibinfo{year}{2007}{\natexlab{a}}).

\bibitem[{\citenamefont{Mitroy and Bromley}(2003)}]{mitroy}
\bibinfo{author}{\bibfnamefont{J.}~\bibnamefont{Mitroy}} \bibnamefont{and}
  \bibinfo{author}{\bibfnamefont{M.~W.~J.} \bibnamefont{Bromley}},
  \bibinfo{journal}{Phys. Rev. A} \textbf{\bibinfo{volume}{68}},
  \bibinfo{pages}{052714} (\bibinfo{year}{2003}).

\bibitem[{\citenamefont{Sahoo and Das}(2008)}]{sahoopra08}
\bibinfo{author}{\bibfnamefont{B.~K.} \bibnamefont{Sahoo}} \bibnamefont{and}
  \bibinfo{author}{\bibfnamefont{B.~P.} \bibnamefont{Das}},
  \bibinfo{journal}{(Submitted to PRA) arXiv:0801.0295}
  (\bibinfo{year}{2008}).

\bibitem[{\citenamefont{Sahoo et~al.}(2007{\natexlab{b}})\citenamefont{Sahoo,
  Das, Chaudhuri, and Mukherjee}}]{sahoojcmse07}
\bibinfo{author}{\bibfnamefont{B.~K.} \bibnamefont{Sahoo}},
  \bibinfo{author}{\bibfnamefont{B.~P.} \bibnamefont{Das}},
  \bibinfo{author}{\bibfnamefont{R.~K.} \bibnamefont{Chaudhuri}},
  \bibnamefont{and}
  \bibinfo{author}{\bibfnamefont{D.}~\bibnamefont{Mukherjee}},
  \bibinfo{journal}{J. Comp. Methods in Sci. and Eng.}
  \textbf{\bibinfo{volume}{7}}, \bibinfo{pages}{57}
  (\bibinfo{year}{2007}{\natexlab{b}}).

\bibitem[{\citenamefont{Pipin and Bishop}(1992)}]{pipinpra92}
\bibinfo{author}{\bibfnamefont{J.}~\bibnamefont{Pipin}} \bibnamefont{and}
  \bibinfo{author}{\bibfnamefont{D.~M.} \bibnamefont{Bishop}},
  \bibinfo{journal}{Phys. Rev. A} \textbf{\bibinfo{volume}{45}},
  \bibinfo{pages}{2736} (\bibinfo{year}{1992}).

\bibitem[{\citenamefont{Themelis and Nicolaides}(1995)}]{themelispra95}
\bibinfo{author}{\bibfnamefont{S.~I.} \bibnamefont{Themelis}} \bibnamefont{and}
  \bibinfo{author}{\bibfnamefont{C.~A.} \bibnamefont{Nicolaides}},
  \bibinfo{journal}{Phys. Rev. A} \textbf{\bibinfo{volume}{51}},
  \bibinfo{pages}{2801} (\bibinfo{year}{1995}).

\bibitem[{\citenamefont{M\'erawa and R\'erat}(1998)}]{merawajcp98}
\bibinfo{author}{\bibfnamefont{M.}~\bibnamefont{M\'erawa}} \bibnamefont{and}
  \bibinfo{author}{\bibfnamefont{M.}~\bibnamefont{R\'erat}},
  \bibinfo{journal}{J. Chem. Phys.} \textbf{\bibinfo{volume}{108}},
  \bibinfo{pages}{7060} (\bibinfo{year}{1998}).

\bibitem[{\citenamefont{Zhang et~al.}(2007)\citenamefont{Zhang, Mitroy, and
  Bromley}}]{zhangpra07}
\bibinfo{author}{\bibfnamefont{J.-Y.} \bibnamefont{Zhang}},
  \bibinfo{author}{\bibfnamefont{J.}~\bibnamefont{Mitroy}}, \bibnamefont{and}
  \bibinfo{author}{\bibfnamefont{M.}~\bibnamefont{Bromley}},
  \bibinfo{journal}{Phys. Rev. A} \textbf{\bibinfo{volume}{75}},
  \bibinfo{pages}{042509} (\bibinfo{year}{2007}).

\bibitem[{\citenamefont{Magnier and Aubert-Frecon}(2002)}]{magnierjqsr02}
\bibinfo{author}{\bibfnamefont{S.}~\bibnamefont{Magnier}} \bibnamefont{and}
  \bibinfo{author}{\bibfnamefont{M.}~\bibnamefont{Aubert-Frecon}},
  \bibinfo{journal}{J. Quant. Spec. Rad. Trans.} \textbf{\bibinfo{volume}{75}},
  \bibinfo{pages}{121} (\bibinfo{year}{2002}).

\bibitem[{\citenamefont{Ashby and van
  Wijngaarden}(2003{\natexlab{a}})}]{ashbyjqsrt02}
\bibinfo{author}{\bibfnamefont{R.}~\bibnamefont{Ashby}} \bibnamefont{and}
  \bibinfo{author}{\bibfnamefont{W.}~\bibnamefont{van Wijngaarden}},
  \bibinfo{journal}{J. Quant. Spec. Rad. Trans.} \textbf{\bibinfo{volume}{76}},
  \bibinfo{pages}{467} (\bibinfo{year}{2003}{\natexlab{a}}).

\bibitem[{\citenamefont{Safronova et~al.}(1999)\citenamefont{Safronova,
  Johnson, and Derevianko}}]{safronovapra99}
\bibinfo{author}{\bibfnamefont{M.~S.} \bibnamefont{Safronova}},
  \bibinfo{author}{\bibfnamefont{W.~R.} \bibnamefont{Johnson}},
  \bibnamefont{and}
  \bibinfo{author}{\bibfnamefont{A.}~\bibnamefont{Derevianko}},
  \bibinfo{journal}{Phys. Rev. A} \textbf{\bibinfo{volume}{60}},
  \bibinfo{pages}{4476} (\bibinfo{year}{1999}).

\bibitem[{\citenamefont{Angel and Sandars}(1968)}]{angel68}
\bibinfo{author}{\bibfnamefont{J.}~\bibnamefont{Angel}} \bibnamefont{and}
  \bibinfo{author}{\bibfnamefont{P.}~\bibnamefont{Sandars}},
  \bibinfo{journal}{Proc. Roy. Soc. A.} \textbf{\bibinfo{volume}{305}},
  \bibinfo{pages}{125} (\bibinfo{year}{1968}).

\bibitem[{\citenamefont{Itano}(2000)}]{itano00}
\bibinfo{author}{\bibfnamefont{W.~M.} \bibnamefont{Itano}},
  \bibinfo{journal}{J. Research NIST} \textbf{\bibinfo{volume}{105}},
  \bibinfo{pages}{829} (\bibinfo{year}{2000}).

\bibitem[{\citenamefont{Blundell et~al.}(1992)\citenamefont{Blundell,
  Sapirstein, and Johnson}}]{blundellprd92}
\bibinfo{author}{\bibfnamefont{S.~A.} \bibnamefont{Blundell}},
  \bibinfo{author}{\bibfnamefont{J.}~\bibnamefont{Sapirstein}},
  \bibnamefont{and} \bibinfo{author}{\bibfnamefont{W.~R.}
  \bibnamefont{Johnson}}, \bibinfo{journal}{Phys. Rev. D}
  \textbf{\bibinfo{volume}{45}}, \bibinfo{pages}{1602} (\bibinfo{year}{1992}).

\bibitem[{\citenamefont{Mukherjee et~al.}(1977)\citenamefont{Mukherjee, Moitra,
  and Mukhopadhyay}}]{mukherjee77}
\bibinfo{author}{\bibfnamefont{D.}~\bibnamefont{Mukherjee}},
  \bibinfo{author}{\bibfnamefont{R.}~\bibnamefont{Moitra}}, \bibnamefont{and}
  \bibinfo{author}{\bibfnamefont{A.}~\bibnamefont{Mukhopadhyay}},
  \bibinfo{journal}{Mol. Physics} \textbf{\bibinfo{volume}{33}},
  \bibinfo{pages}{955} (\bibinfo{year}{1977}).

\bibitem[{\citenamefont{Lindgren}(1978)}]{lindgren78}
\bibinfo{author}{\bibfnamefont{I.}~\bibnamefont{Lindgren}}, in
  \emph{\bibinfo{booktitle}{Atomic, molecular, and solid-state theory,
  collision phenomena, and computational methods}}, edited by
  \bibinfo{editor}{\bibfnamefont{P.-O.} \bibnamefont{Iwdin}} \bibnamefont{and}
  \bibinfo{editor}{\bibfnamefont{Y.}~\bibnamefont{Ahrn}}
  (\bibinfo{organization}{International Journal of Quantum Chemistry, Quantum
  Chemistry Symposium}, \bibinfo{year}{1978}), vol.~\bibinfo{volume}{12},
  p.~\bibinfo{pages}{33}.

\bibitem[{\citenamefont{Lindgen and Morrison}(1985)}]{lindgren85}
\bibinfo{author}{\bibfnamefont{I.}~\bibnamefont{Lindgen}} \bibnamefont{and}
  \bibinfo{author}{\bibfnamefont{J.}~\bibnamefont{Morrison}},
  \emph{\bibinfo{title}{Atomic Many-Body Theory}}
  (\bibinfo{publisher}{Springer-Verlag, Berlin}, \bibinfo{year}{1985}).

\bibitem[{\citenamefont{Kaldor}(1987)}]{kaldorjcp87}
\bibinfo{author}{\bibfnamefont{U.}~\bibnamefont{Kaldor}}, \bibinfo{journal}{J.
  Chem. Phys.} \textbf{\bibinfo{volume}{87}}, \bibinfo{pages}{4693}
  (\bibinfo{year}{1987}).

\bibitem[{\citenamefont{Molof et~al.}(1974)\citenamefont{Molof, Schwartz,
  Miller, and Bederson}}]{molof}
\bibinfo{author}{\bibfnamefont{R.~W.} \bibnamefont{Molof}},
  \bibinfo{author}{\bibfnamefont{H.~L.} \bibnamefont{Schwartz}},
  \bibinfo{author}{\bibfnamefont{T.~M.} \bibnamefont{Miller}},
  \bibnamefont{and} \bibinfo{author}{\bibfnamefont{B.}~\bibnamefont{Bederson}},
  \bibinfo{journal}{Phys. Rev. A} \textbf{\bibinfo{volume}{10}},
  \bibinfo{pages}{1131} (\bibinfo{year}{1974}).

\bibitem[{\citenamefont{Miffre et~al.}(2006)\citenamefont{Miffre, Jacquery,
  B{\"u}chner, Trenec, and Vigue}}]{miffre}
\bibinfo{author}{\bibfnamefont{A.}~\bibnamefont{Miffre}},
  \bibinfo{author}{\bibfnamefont{M.}~\bibnamefont{Jacquery}},
  \bibinfo{author}{\bibfnamefont{M.}~\bibnamefont{B{\"u}chner}},
  \bibinfo{author}{\bibfnamefont{G.}~\bibnamefont{Trenec}}, \bibnamefont{and}
  \bibinfo{author}{\bibfnamefont{J.}~\bibnamefont{Vigue}},
  \bibinfo{journal}{Phys. Rev. A} \textbf{\bibinfo{volume}{73}},
  \bibinfo{pages}{011603(R)} (\bibinfo{year}{2006}).

\bibitem[{\citenamefont{Ashby and van
  Wijngaarden}(2003{\natexlab{b}})}]{ashbyjqsr03}
\bibinfo{author}{\bibfnamefont{R.}~\bibnamefont{Ashby}} \bibnamefont{and}
  \bibinfo{author}{\bibfnamefont{W.~A.} \bibnamefont{van Wijngaarden}},
  \bibinfo{journal}{J. Quant. Spec. Rad. Trans.} \textbf{\bibinfo{volume}{76}},
  \bibinfo{pages}{467} (\bibinfo{year}{2003}{\natexlab{b}}).

\bibitem[{\citenamefont{Windholz et~al.}(1992)\citenamefont{Windholz, Musso,
  Zerza, and Jager}}]{windholzpra02}
\bibinfo{author}{\bibfnamefont{L.}~\bibnamefont{Windholz}},
  \bibinfo{author}{\bibfnamefont{M.}~\bibnamefont{Musso}},
  \bibinfo{author}{\bibfnamefont{G.}~\bibnamefont{Zerza}}, \bibnamefont{and}
  \bibinfo{author}{\bibfnamefont{H.}~\bibnamefont{Jager}},
  \bibinfo{journal}{Phys. Rev. A} \textbf{\bibinfo{volume}{46}},
  \bibinfo{pages}{5812} (\bibinfo{year}{1992}).

\bibitem[{\citenamefont{Ashby et~al.}(2003)\citenamefont{Ashby, Clarke, and van
  Wijngaarden}}]{ashbyepjd03}
\bibinfo{author}{\bibfnamefont{R.}~\bibnamefont{Ashby}},
  \bibinfo{author}{\bibfnamefont{J.~J.} \bibnamefont{Clarke}},
  \bibnamefont{and} \bibinfo{author}{\bibfnamefont{W.~A.} \bibnamefont{van
  Wijngaarden}}, \bibinfo{journal}{Eur. Phys. J. D}
  \textbf{\bibinfo{volume}{23}}, \bibinfo{pages}{327} (\bibinfo{year}{2003}).

\bibitem[{\citenamefont{Hunter et~al.}(1991)\citenamefont{Hunter, Jr.,
  Berkeland, and Boshier}}]{hunterpra91}
\bibinfo{author}{\bibfnamefont{L.~R.} \bibnamefont{Hunter}},
  \bibinfo{author}{\bibfnamefont{D.~K.} \bibnamefont{Jr.}},
  \bibinfo{author}{\bibfnamefont{D.~J.} \bibnamefont{Berkeland}},
  \bibnamefont{and} \bibinfo{author}{\bibfnamefont{M.~G.}
  \bibnamefont{Boshier}}, \bibinfo{journal}{Phys. Rev. A}
  \textbf{\bibinfo{volume}{44}}, \bibinfo{pages}{6140} (\bibinfo{year}{1991}).

\bibitem[{\citenamefont{Parpia et~al.}()\citenamefont{Parpia, {Froese Fischer},
  and Grant}}]{parpia}
\bibinfo{author}{\bibfnamefont{F.}~\bibnamefont{Parpia}},
  \bibinfo{author}{\bibfnamefont{C.}~\bibnamefont{{Froese Fischer}}},
  \bibnamefont{and} \bibinfo{author}{\bibfnamefont{I.~P.} \bibnamefont{Grant}},
  \bibinfo{note}{unpublished}.

\bibitem[{\citenamefont{Chaudhuri et~al.}(1999)\citenamefont{Chaudhuri, Panda,
  and Das}}]{chaudhuripra99}
\bibinfo{author}{\bibfnamefont{R.~K.} \bibnamefont{Chaudhuri}},
  \bibinfo{author}{\bibfnamefont{P.~K.} \bibnamefont{Panda}}, \bibnamefont{and}
  \bibinfo{author}{\bibfnamefont{B.~P.} \bibnamefont{Das}},
  \bibinfo{journal}{Phys. Rev. A} \textbf{\bibinfo{volume}{59}},
  \bibinfo{pages}{1187} (\bibinfo{year}{1999}).

\bibitem[{\citenamefont{Majumder et~al.}(2001)\citenamefont{Majumder, Geetha,
  Merlitz, and Das}}]{majumderjpb01}
\bibinfo{author}{\bibfnamefont{S.}~\bibnamefont{Majumder}},
  \bibinfo{author}{\bibfnamefont{K.}~\bibnamefont{Geetha}},
  \bibinfo{author}{\bibfnamefont{H.}~\bibnamefont{Merlitz}}, \bibnamefont{and}
  \bibinfo{author}{\bibfnamefont{B.}~\bibnamefont{Das}}, \bibinfo{journal}{J.
  Phys. B} \textbf{\bibinfo{volume}{34}}, \bibinfo{pages}{2841}
  (\bibinfo{year}{2001}).

\bibitem[{\citenamefont{Marinescu et~al.}(1994)\citenamefont{Marinescu,
  Sadeghpour, and Dalgarno}}]{marinescu}
\bibinfo{author}{\bibfnamefont{M.}~\bibnamefont{Marinescu}},
  \bibinfo{author}{\bibfnamefont{H.~R.} \bibnamefont{Sadeghpour}},
  \bibnamefont{and} \bibinfo{author}{\bibfnamefont{A.}~\bibnamefont{Dalgarno}},
  \bibinfo{journal}{Phys. Rev. A} \textbf{\bibinfo{volume}{49}},
  \bibinfo{pages}{982} (\bibinfo{year}{1994}).

\bibitem[{\citenamefont{M\'erawa et~al.}(1994)\citenamefont{M\'erawa, R\'erat,
  and Pouchan}}]{merawa94}
\bibinfo{author}{\bibfnamefont{M.}~\bibnamefont{M\'erawa}},
  \bibinfo{author}{\bibfnamefont{M.}~\bibnamefont{R\'erat}}, \bibnamefont{and}
  \bibinfo{author}{\bibfnamefont{C.}~\bibnamefont{Pouchan}},
  \bibinfo{journal}{Phys. Rev. A} \textbf{\bibinfo{volume}{49}},
  \bibinfo{pages}{2493} (\bibinfo{year}{1994}).

\bibitem[{\citenamefont{Patil and Tang}(1997)}]{patil97}
\bibinfo{author}{\bibfnamefont{S.~H.} \bibnamefont{Patil}} \bibnamefont{and}
  \bibinfo{author}{\bibfnamefont{K.~T.} \bibnamefont{Tang}},
  \bibinfo{journal}{J. Chem. Phys.} \textbf{\bibinfo{volume}{106}},
  \bibinfo{pages}{2298} (\bibinfo{year}{1997}).

\bibitem[{\citenamefont{Patil and Tang}(1999)}]{patil99}
\bibinfo{author}{\bibfnamefont{S.~H.} \bibnamefont{Patil}} \bibnamefont{and}
  \bibinfo{author}{\bibfnamefont{K.~T.} \bibnamefont{Tang}},
  \bibinfo{journal}{Chem. Phys. Lett.} \textbf{\bibinfo{volume}{301}},
  \bibinfo{pages}{64} (\bibinfo{year}{1999}).

\bibitem[{\citenamefont{Snow et~al.}(2005)\citenamefont{Snow, Gearba, Komara,
  Lundeen, and Sturrus}}]{snow05}
\bibinfo{author}{\bibfnamefont{E.~L.} \bibnamefont{Snow}},
  \bibinfo{author}{\bibfnamefont{M.~A.} \bibnamefont{Gearba}},
  \bibinfo{author}{\bibfnamefont{R.~A.} \bibnamefont{Komara}},
  \bibinfo{author}{\bibfnamefont{S.~R.} \bibnamefont{Lundeen}},
  \bibnamefont{and} \bibinfo{author}{\bibfnamefont{W.~G.}
  \bibnamefont{Sturrus}}, \bibinfo{journal}{Phys. Rev. A}
  \textbf{\bibinfo{volume}{71}}, \bibinfo{pages}{022510}
  (\bibinfo{year}{2005}).

\bibitem[{\citenamefont{Porsev and Derevianko}(2003)}]{porsev03}
\bibinfo{author}{\bibfnamefont{S.~G.} \bibnamefont{Porsev}} \bibnamefont{and}
  \bibinfo{author}{\bibfnamefont{A.}~\bibnamefont{Derevianko}},
  \bibinfo{journal}{J. Chem. Phys.} \textbf{\bibinfo{volume}{119}},
  \bibinfo{pages}{844} (\bibinfo{year}{2003}).

\end{thebibliography}


\end{document}